# Single crystal growth and structural characterization of FeTe$_{1-x}$S$_x$

Yoshikazu Mizuguchi[1,2,3], Keita Deguchi[1,2,3], Toshinori Ozaki[1,2], Masanori Nagao[1,4], Shunsuke Tsuda[1,2], Takahide Yamaguchi[1,2] and Yoshihiko Takano[1,2,3]

*Abstract*— We have synthesized single crystals of FeTe$_{1-x}$S$_x$ and investigated the structural and physical properties. The actual S concentration was found to be significantly smaller than the starting nominal value. The solubility limit of S for the Te site for the crystals grown using a self-flux method was estimated to be around 12 %. Although a suppression of antiferromagnetic transition was observed by S substitution, bulk superconductivity was not observed. A phase diagram of as-grown FeTe$_{1-x}$S$_x$ single crystal was established on the basis of the structural and physical properties.

## I. INTRODUCTION

Fe chalcogenide with a PbO-type crystal structure is the simplest Fe-based superconductor [1-4]. Due to its simple structure composed of only Fe-chalcogen layers, Fe chalcogenide is an ideal system for understanding of the mechanism of Fe-based superconductivity. FeSe is a superconductor with a transition temperature $T_c$ = 8 K, and shows a dramatic enhancement of $T_c$ up to 37 K under high pressures around 4-6 GPa [3,5-8]. The enhancement of $T_c$ under high pressure seems to be related to changes in Se height from the Fe layer and/or an enhancement of the antiferromagnetic fluctuations [6,8,9-11].

Whereas FeSe is superconducting, FeTe exhibits an antiferromagnetic ordering below 70 K while it has a crystal structure analogous to superconducting FeSe. However, the magnetic ordering is suppressed by a partial substitution of S or Se for Te and superconductivity appears [12-15]. It is the intrinsic properties of the Fe-Te compounds which should shed light on the mechanism and the evolution of superconductivity in the Fe-Te compounds.

The synthesis of FeTe$_{1-x}$S$_x$ superconductor is difficult due to a large difference of ionic radius between S$^{2-}$ (186 pm) and Te$^{2-}$ (211 pm). Coexistence of superconducting and antiferromagnetic phase was observed in FeTe$_{1-x}$S$_x$ single crystals [16], which indicates that the solubility limit of S for the Te site is low. Recently, we have reported moisture-induced and oxygen-annealing-induced superconductivity in FeTe$_{0.8}$S$_{0.2}$, which was synthesized using solid-state reaction and did not show bulk superconductivity just after the synthesis [17,18]. In both cases, the shrinkage of lattice was observed for the superconducting samples. To elucidate why superconductivity is induced by the lattice shrinkage, we synthesized single crystals of FeTe$_{1-x}$S$_x$ with several nominal compositions. Here we discuss the structural and physical properties.

## II. EXPERIMENTAL DETAILS

Single crystals of FeTe$_{1-x}$S$_x$ were synthesized using a self-flux method. Fe powder (99.9 %), S grains (99% up) and Te grains (99.999 %) with nominal compositions ($x$ = 0.05, 0.1, 0.15, 0.2, 0.25, 0.3, 0.4 and 0.5) were placed in an alumina crucible and sealed into an evacuated quartz tube. The samples were heated up to 1050 ºC in 10 hours, kept for 20 hours and cooled down to 650 ºC with a rate of –4 ºC/hour.

The actual composition of the crystal was determined using electron probe micro analyzer (EPMA). The value was estimated using an average of four measurements. The standard deviation was used as an error bar. The lattice constant $c$ for the obtained crystals was estimated by x-ray diffraction using the $2\theta$-$\theta$ method with Cu-K$\alpha$ radiation. The x-ray profile from $2\theta$ = 5 to 70 deg. was collected using Mini Flex (Rigaku). Temperature dependence of resistivity from 300 to 2 K was measured using the four-terminal method.

## III. RESULTS AND DISCUSSION

Plate-like single crystals were obtained for all nominal compositions. Photographs of the obtained crystals with $x$ = 0.05 and 0.1 are shown in Fig. (a) and (b), respectively. The color of those crystals is silver and shining. With increasing S concentration, the size of the obtained single crystals tended to be smaller.

The actual S concentration determined by EPMA ($x_E$) is summarized in Table 1, and plotted in Fig. 2 as a function of nominal $x$. The estimated $x_E$ is found to be smaller than nominal $x$. Although $x_E$ increases with increasing nominal $x$ up to $x$ = 0.2, it saturates above $x$ = 0.2. This indicates that the solubility limit

Manuscript received 3 August 2010.
This work was partly supported by a Grant-in-Aid for Scientific Research (KAKENHI).
Y. Mizuguchi, K. Deguchi, T. Ozaki, M. Nagao, S. Tsuda, T. Yamaguchi and Y. Takano are with National Institute for Materials Science, 1-2-1, Sengen, Tsukuba, 305-0047, Japan (corresponding author to provide phone: +81-29-859-2842; fax: +81-29-859-2601; e-mail: MIZUGUCHI.Yoshikazu@nims.go.jp).
Y. Mizuguchi, K. Deguchi, T. Ozaki, S. Tsuda, T. Yamaguchi and Y. Takano are with JST-TRIP (Transformative Research Project on Iron Pnictides), 1-2-1, Sengen, Tsukuba, 305-0047, Japan.
Y. Mizuguchi, K. Deguchi and Y. Takano are with University of Tsukuba, 1-1-1 Tennodai, Tsukuba, 305-8571, Japan.
M. Nagao is with University of Yamanashi, 7-32 Miyamae, Kofu, 400-8511, Japan.

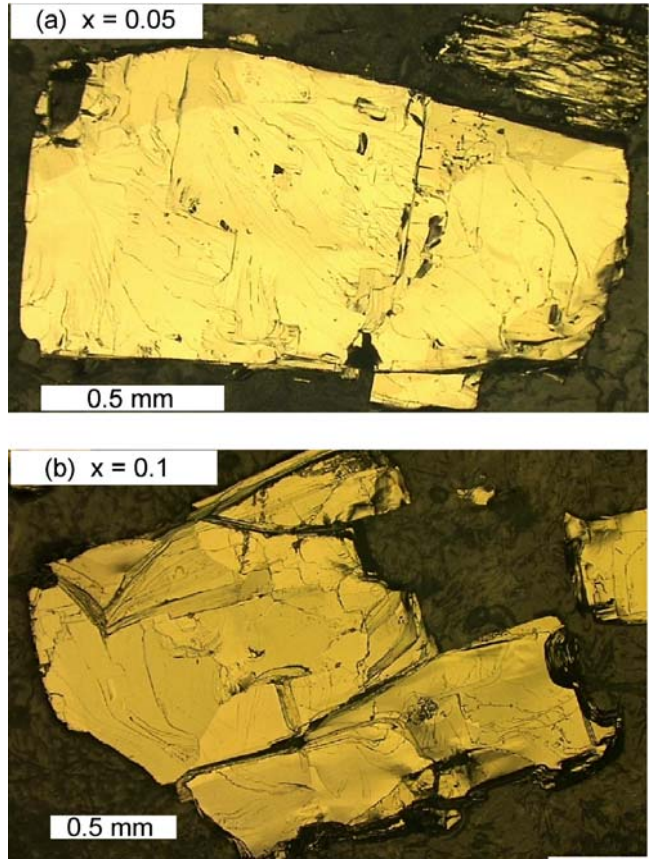

Fig. 1. Optical microscope images for $x = 0.05$ (a) and 0.1 (b).

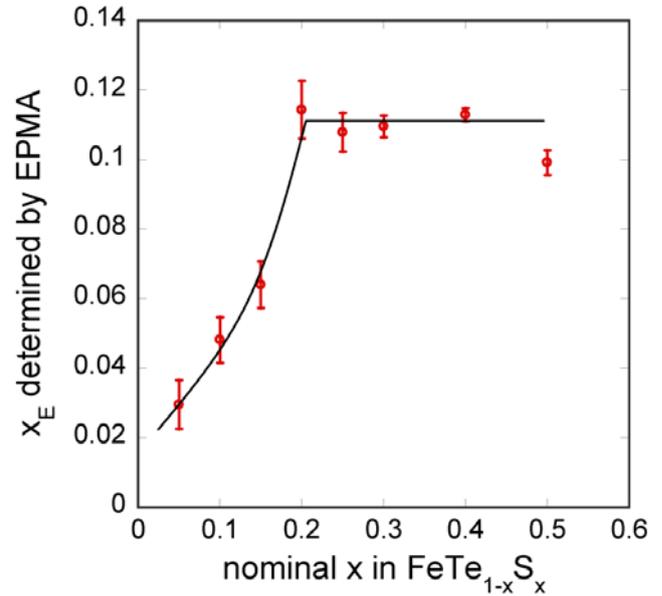

Fig. 2. Nominal $x$ dependence of $x_E$ determined by EPMA.

of S for Te for the single crystals grown using the self-flux method seems to be around 12 %. Here we have estimated a ratio of S/Te, because a comparably large dispersion was observed for Fe concentration. The value of $1+d$ in $Fe_{1+d}Te_{1-x}S_x$ was calculated to be 0.98 - 1.05.

Figure 3(a) shows the typical x-ray diffraction patterns collected using several single crystals. Only the peaks corresponding to (00$l$) reflection were observed. An enlargement of the 002 peak is shown in Fig. 3(b). The peak clearly shifts to higher angles with increasing S concentration. The lattice constant $c$ calculated using these peak positions is summarized in Table 1 and plotted in Fig. 4 as a function of starting nominal $x$. The $c$ value decreases with increasing nominal $x$ up to $x = 0.25$, and saturates with $x > 0.25$, which is consistent with the result from EPMA. For comparison, $x_E$ dependence of $c$ values for $FeTe_{1-x}S_x$ are plotted with lattice constants of $FeTe_{1-x}Se_x$ in the inset of Fig. 4. The slope of decrease in $c$-axis for $FeTe_{1-x}S_x$ almost corresponds to that of $FeTe_{1-x}Se_x$. This suggests that both $x_E$ estimated by EPMA and lattice constant $c$ estimated from x-ray diffraction are reliable, because the ionic radius of $S^{2-}$ (186 pm) is close to that of $Se^{2-}$ (191 pm).

| $x$ (nominal) | $c$ (Å)    | $x$ (EPMA) |
|---------------|------------|------------|
| 0.05          | 6.2798(9)  | 0.030(7)   |
| 0.1           | 6.265(1)   | 0.048(7)   |
| 0.15          | 6.250 (1)  | 0.064(7)   |
| 0.2           | 6.245(1)   | 0.114(8)   |
| 0.25          | 6.239(1)   | 0.108(6)   |
| 0.3           | 6.241(2)   | 0.110(3)   |
| 0.4           | 6.243(3)   | 0.113(2)   |
| 0.5           | 6.245 (3)  | 0.099(4)   |

Table 1. Lattice constant $c$ as determined by x-ray diffraction and S concentration $x$ as determined by EPMA.

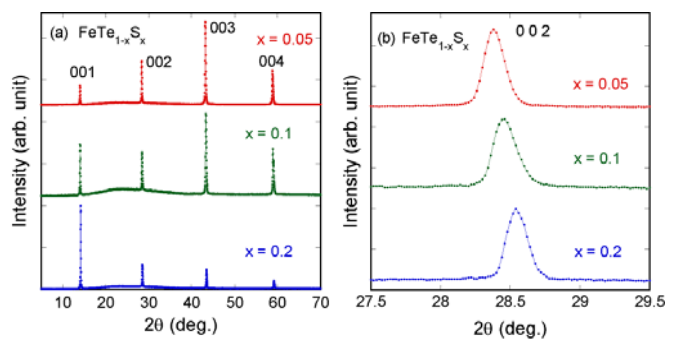

Fig. 3. (a)Normalized x-ray diffraction profiles for $x = 0.05$, 0.1 and 0.2. (b)An enlargement of 002 reflections.

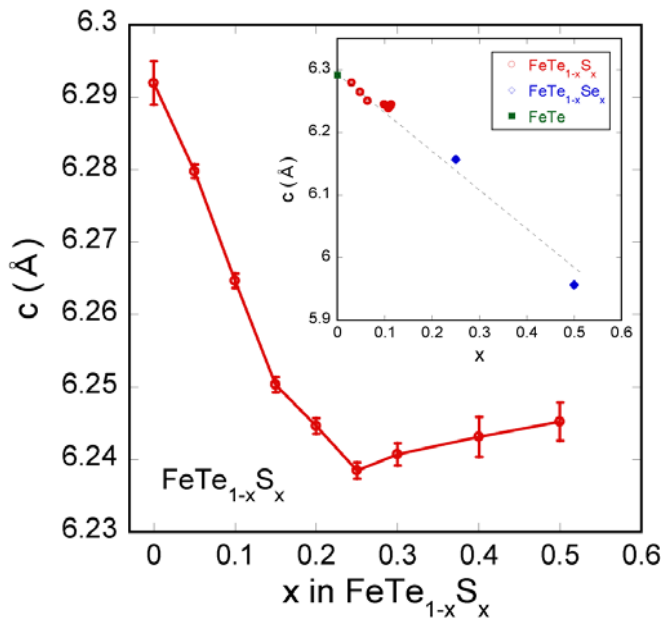

Fig. 4. Nominal $x$ dependence of lattice constant $c$ of the obtained single crystals. The inset shows a comparison of shrinkage of $c$ axis between FeTe$_{1-x}$S$_x$ and FeTe$_{1-x}$Se$_x$.

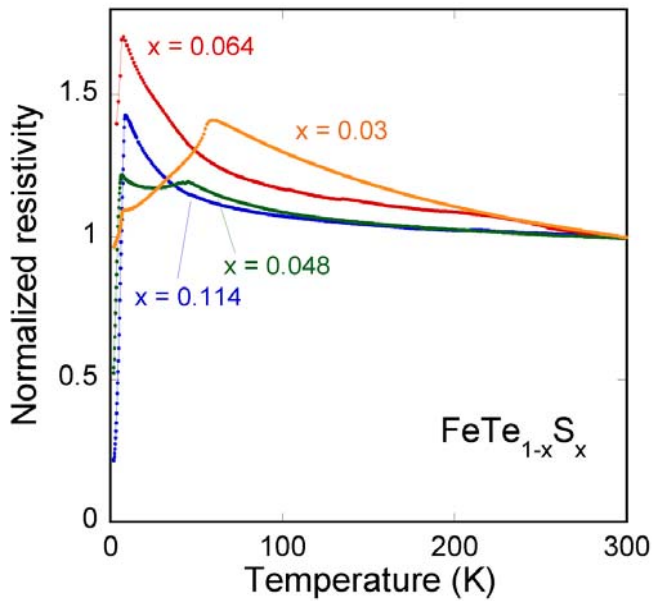

Fig. 5. Temperature dependence of resistivity for $x = 0.03$, 0.048, 0.064 and 0.114.

Figure 5 shows the temperature dependence of resistivity for $x = 0.03 – 0.114$, where the resistivity is normalized at 300 K for comparison. The anomaly corresponding to the antiferromagnetic and structural transition observed in FeTe is suppressed by S substitution. For $x = 0.03$ and 0.048, the transition temperature is estimated to be 58 K and 45 K, respectively. The transition disappears above $x = 0.064$. In spite of the suppression of the magnetic transition, spurious superconductivity was observed for $x = 0.064$ and 0.114. This suggests that the 11.4 % substitution of S for Te is insufficient to achieve bulk superconductivity. To achieve superconductivity in these single crystals, lattice would have to be compressed by an increase of S concentration or post-treatments such as air exposure or oxygen annealing, which compress the lattice and induce bulk superconductivity as found in polycrystalline FeTe$_{1-x}$S$_x$.

On the basis of the investigation of the structural and physical properties for FeTe$_{1-x}$S$_x$ single crystals, phase diagram is summarized as shown in Fig. 6. Red circles indicate structural and antiferromagnetic transition temperature $T_s$ determined from resistivity measurement. The S concentration in Fig. 6 is the value determined by EPMA. The scenario that S substitution at the Te site suppresses the antiferromagnetic ordering and induces superconductivity is the same as that for the FeTe$_{1-x}$Se$_x$ superconductor. However, bulk superconductivity is not observed for the as-grown single crystals of FeTe$_{1-x}$S$_x$ grown using the self-flux method. The lack of superconductivity could be due to the low S concentration and/or the existence of large amount of excess Fe, which has a magnetic moment and interferes with superconductivity. The amount of the excess Fe in Fe$_{1+d}$Te tends to decrease with S or Se substitution for Te [19-22]. Due to the low solubility limit of S for the Te site in the FeTe$_{1-x}$S$_x$ single crystals, large amount of excess Fe may exist and interfere with bulk superconductivity. To clarify the factors important for superconductivity in Fe chalcogenide, search for a post treatment that induces bulk superconductivity in the FeTe$_{1-x}$S$_x$ single crystals should be performed.

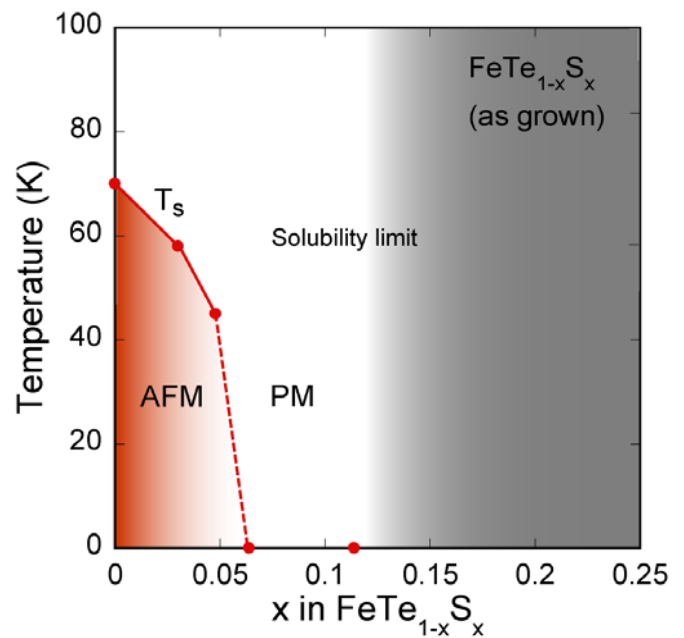

Fig. 6. Phase diagram of as-grown single crystal of FeTe$_{1-x}$S$_x$. Red filled circles are structural and antiferromagnetic transition temperature $T_s$. AFM and PM stand for antiferromagnetism and paramagnetism, respectively.

## IV. CONCLUSION

We synthesized single crystals of FeTe$_{1-x}$S$_x$ with nominal compositions of $x = 0 - 0.5$ using the self-flux method. From both x-ray diffraction and EPMA, the solubility limit of S for the Te site for the crystals grown using the self-flux method is found to be about 12%. The antiferromagnetic transition, observed at 70 K in FeTe, is suppressed with increasing S concentration, and disappears above $x \sim 0.06$. Although long-range magnetic ordering is suppressed for $x > 0.06$, superconductivity is not observed. The lack of superconductivity could be due to the low S concentration and/or the existence of excess Fe, which interferes the appearance of bulk superconductivity in FeTe-based compounds.